# Implementing augmented reality technology to measure structural changes across time

Jiaqi Xu[1], Elijah Wyckoff[2], Marlon Aguero[1], John-Wesley Hanson[1], Fernando Moreu[1]*, Derek Doyle[3]

*1 Department of Civil, Construction & Environmental Engineering, University of New Mexico, Albuquerque, NM, USA*

*2 Department of Mechanical Engineering, University of New Mexico, Albuquerque, NM, USA*

*3 Air Force Research Laboratory, Space Vehicles Directorate, Kirtland Air Force Base, Albuquerque, NM, USA*

In recent years, augmented reality (AR) technology has been increasingly employed in structural health monitoring (SHM). In the case of conditions following a seismic event, inspections are conducted to evaluate the progression of the damage pattern quantitatively and efficiently respond if the displacement pattern is determined to be unsafe. Additionally, quantification of nearby structural changes over short-term and long-term periods can provide building inspectors with information to improve safety. This paper proposes the Time Machine Measure (TMM) application on an Augmented Reality (AR) Head-Mounted-Device (HMD) platform. The main function of the TMM application is to restore the saved meshes of a past environment and overlay them onto the real environment so that inspectors can intuitively measure structural deformation and other movement across time. The proposed TMM application was verified by experiments meant to simulate a real-world inspection.

## 1 INTRODUCTION

Built environment is defined as human-made surroundings designed for human activities (Handy et al. 2002). Various emergencies and hazards can happen in a built environment, e.g., earthquakes. Ensuring the safety of inspectors is a priority, and hazards exist that threaten their safety. Seismic hazards include mainshock and secondary hazards, e.g., aftershocks. Aftershocks can suddenly happen after the mainshock in days, weeks, months, or even years. For example, the Mw7.8 Gorkha earthquake struck central Nepal on April 25th, 2015, followed by an Mw6.7 aftershock on April 26th (1 day after the mainshock) and an Mw7.3 aftershock on May 12th (17 days after the mainshock) (Arora et al. 2017). Although the magnitude of aftershocks is smaller than the mainshock, the risk of structural deformation during aftershocks is higher due to the dynamic characteristic change of the damaged structures.

Earthquake rescuers agree that the first 72 hours are essential for life-saving, usually known as the 'golden relief time' (Ochoa et al. 2007; Dourandish, Zumel, and Manno 2007; Fiedrich, Gehbauer, and Rickers 2000). However, the first 72-hour entry for earthquake inspectors has the maximum probability of structural movement caused structural collapse (Felzer, Abercrombie, and Ekström 2003). Therefore, reliable and easy-to-operate observation methods for monitoring environmental activities over time are vital (Gallagher, Reasenberg, and Poland 1999). ATC-20 makes a general recommendation about the maximum waiting time for emergency access by considering earthquake mainshock magnitude and days after the mainshock. The limitation of ATC-20 is that the officers make the recommendations based on statistical analyses and experiences (Applied Technology Council (ATC) 1989; ATC 2005). Hence ATC-20 cannot provide recommendations based on real-time movement that varies with each disaster and building/structure.

There are three major implementations of AR technology in emergency management: i) simulating disaster environments; ii) studying disaster factors; and iii) training (Zhu and Li 2021). For simulating disaster environments, researchers have implemented AR technology to draw AR maps of the building so first responders can detect rooms that may be invisible due to debris blockage or smoke during rescue operations (Yang et al. 2018; Sebillo et al. 2016; Sharma, Stigal, and Bodempudi 2020). An AR map can also help inspectors make responsible decisions with planning routes to avoid tie-ups or unsafe regions. AR has also been implemented to simulate building collapse procedures in an earthquake (Leebmann 2004; Tadokoro et al. 2000; Kitano et al. 1999). Researchers have been focused on AR-aided rapid post-disaster building assessment to provide more valuable information support for inspectors. With AR technology, inspectors can conduct a fast building assessment before entering the damaged building. Kamat and El-Tawil (2007; 2005) developed an HMD to automatically compare the CAD images (i.e., original design) with the current state of the damaged building. The structural damage can be detected by measuring the critical differences between the baseline image and the facility's realistic view.

The concept of overlapping previously saved virtual images onto real environments is referred to as 'time machine' in this paper. Zoellner et al. (2008) developed a time machine application, Reality Filtering, to visualize paintings of buildings and frescos seamlessly superimposed onto cultural heritage and architecture area. Holmgren, Johansson, and Andersson (2014) developed another time machine software application implementing an AR view to access multimedia information of the past and the present when standing in a specific position. These applications of time machine can only restore pre-input images, i.e., they are not reference-free. Also, current time machine applications cannot measure movement of structures and objects.

This paper summarizes the design, development, and validation of Time Machine Measure, a novel tool to measure and share reference-free changes during inspection. TMM can capture and measure changes in environmental movements and position over time and can assist inspectors by informing them of these changes over short and extended time periods. Additionally, the TMM application is contact-free and solely controlled by gestures, freeing both hands for other inspection tasks. Voice input can be used with the application as well, allowing the user perform actions in the AR application with their gaze and voice. No prerequisite for implementing the pre-input of the CAD model is needed in the TMM application. In the following sections, the authors first design theory and summarize the software framework design, presenting both the human-device interface and operational instructions.

## 2 MEASUREMENT DURING INSPECTIONS

Measurement across time can be of value during post-seismic event inspections. Damage patterns have varying levels of seismic vulnerability, and the various damage patterns observed in built environments are classified by Okada and Takai (1999). For example, structural deformation consisting of inter-story drift and rigid body rotation (X. Lu et al. 2017; Ghobarah 2004) can indicate progressive decay on the structure's stability. It would be valuable to notice structural deformation across time and other local movements, for example inspection of a residential building post-earthquake. This information can inform inspectors of real-time and short-term changes in their surroundings and can also help them classify damage and monitor changes over an extended time period.

Today's solutions to measure structural deformation during the inspection process include comparing the current building visually with the embedded blueprint and recognize the discrepancies (Kamat and El Tawil 2007; Kamat and El-Tawil 2005). Okada and Takai (1999) cite the importance of hazard reduction by identifying damage to individual structures in post-earthquake inspections. Under the premise of enough time and access to the design blueprint, today's solutions work well for inspectors to quickly conduct a building safety assessment if they plan to enter a structure. However, time and design blueprints are not as useful for multiple short-term assessments.

Additionally, today's solutions do not enable direct observation of the structural changes across time. The included initial construction error overestimates the risk level of damaged buildings and shortens the precious rescue time. Figure 1 illustrates a straightforward example of the value of measuring structural deformation after a seismic event. In Figure 1, $d_1$ represents the deformation before the disaster, including the initial construction error and rigid body rotation, which are informative but do not capture changes as time progresses. Conversely, $d_2$ represents the changes that can occur after the disaster. Current devices can only measure $d_1$ while a measuring method of $d_2$ is still not available today.

The authors design two primary functions to fulfill the core requirement of direct measurement of $d_2$: (i) time machine and (ii) measure. The time machine function enables inspectors to recreate structures recorded as a 3D mesh and overlay the past mesh in the present to track deformation across time. The measure function allows inspectors to measure distance between past and present surroundings, i.e., measurement across time.

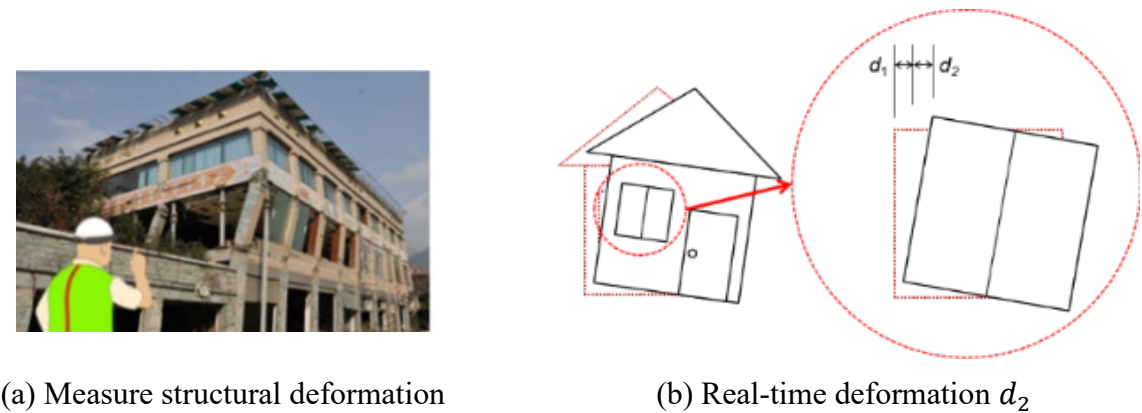

(a) Measure structural deformation              (b) Real-time deformation $d_2$

**Figure 1** Schematic diagram of the structural deformation and inspection

# 3 AR DEVICE SELECTION

The factors to be considered in selecting any AR platform consist of five fundamental categories: general properties, sensors, computational capabilities, connectivity, and display capabilities. Xu, Doyle, and Moreu (2020) summarized the parameters in each category. General properties include release date, country, weight, price, product durability (waterproof, working temperature, relative humidity, and drop safe distance). Sensors consist of camera resolution, head-tracking sensor, eye-tracking sensor, depth sensor, gyroscope, accelerometer, magnetometer, and location sensor (GPS/GLONASS/GALILEO). Computational capabilities include the processing unit, Random Access Memory (RAM), and storage. Connectivity capabilities include WiFi, Bluetooth, USB, and battery capacity. Display capabilities include the field of view (FoV), Optics method, resolution, and refresh rate. An evaluation system can simplify the selection procedure of the AR platform. The composite score of an AR device, $S$, can be described as

$$S = v_i \sum_{i=1}^{m} w_i P_i \tag{1}$$

where $P_i$ represents the $i^{\text{th}}$ parameter considered in the selection. $P_i$ can be a full set or subset of all the parameters illustrated by Xu, Doyle, and Moreu (2020). For a specific implementation, researchers can always add extra parameters as demanded. $w_i$ is the weighing coefficient of $P_i$, measuring the importance of the $i^{\text{th}}$ parameter. $v_i$ is the correction coefficient with values of $\{0 \quad 1\}$. If $P_i$ is a necessary parameter for a specific application, then $v_i = 1$ for the devices with this character, otherwise $v_i = 0$.

The authors select the AR platform for inspection from the sixteen AR devices published in the last three years, with subset parameters provided by Xu, Doyle, and Moreu (2020). According to this research, the weight and price of AR devices will decrease while increasing their outdoor durability, further augmenting its likelihood for field implementations. The authors use a correction coefficient for the depth sensor to meet the time machine function requirement and the rescue requirements. The weighting coefficient $w_i$ of the device with the best behavior in this parameter is 1, and $w_i$ of the other devices are multiplied with a fraction of $P_i/P_{\text{best}}$. For example, the Toshiba dynaEdge AR100 has the maximum working temperature of 60°C, therefore, $w_{\text{temp}}(\text{Toshiba}) = 1$. The maximum working temperature of Google Glass Enterprise Edition 2 is 45°C, therefore, $w_{\text{temp}}(\text{Google}) = 45/60 = 0.75$.

Table 1 summarizes the parameters and the corresponding weighting coefficients to the parameters. The last column of Table 1 lists the total evaluation score of each AR device in descending order. The authors selected the Microsoft HoloLens 2 platform to design the TMM application. Besides the highest score, the Microsoft HoloLens 2 is contact-free and controlled by gestures as an HMD, freeing both hands for other inspection tasks.

# 4 SOFTWARE DESIGN

The aforementioned primary functions of TMM are time machine and measure. The time machine function enables inspectors to observe and compare the current environment to virtual environments from the past. The measure function enables the measurement of deformations over time. The following sections describe in detail the development and use of these two functions.

**Table 1** Evaluation score of AR devices for the emergency rescue

| # | Device | Depth Sensor $v_i = 1\|0$ | Working Temperature $w_i = 1.0$ | Camera 1.0 | RAM 0.8 | Storage 0.8 | Battery 1.0 | FoV 1.0 | Weight 0.5 | Price 0.5 | S |
|---|--------|--------------|-------------------|--------|-----|---------|---------|-----|--------|-------|---|
| 1 | Microsoft HoloLens 2 (2019) | 1 | 0.45 | 0.50 | 0.50 | 0.50 | 1.00 | 0.65 | 0.08 | 0.14 | 3.51 |
| 2 | Magic Leap 1 (2018) | 1 | 0.42 | 0.31 | 1.00 | 1.00 | 0.01 | 0.63 | 0.15 | 0.22 | 3.14 |
| 3 | ThirdEye Gen X2 Mixed Reality Smart Glasses (2019) | 1 | 0.42 | 0.81 | 0.50 | 0.50 | 0.11 | 0.53 | 0.17 | 0.26 | 2.87 |
| 4 | Toshiba dynaEdge AR100 Head Mounted Display (2018) | 1 | 1.00 | 0.31 | 0.00 | 0.00 | 0.06 | 0.59 | 0.98 | 0.17 | 2.54 |
| 5 | DAQRI Smart Glasses (2017) | 1 | 0.42 | 0.50 | 0.13 | 0.50 | 0.35 | 0.38 | 0.14 | 0.10 | 2.26 |
| 6 | Epson Moverio Pro BT-2200 Smart Headset (2017) | 1 | 0.67 | 0.31 | 0.13 | 0.06 | 0.08 | 0.29 | 0.15 | 0.25 | 1.69 |
| 7 | Vuzix M400 Version 1.1.4 (2020) | 0 | 0.75 | 0.50 | 0.75 | 0.50 | 0.01 | 0.21 | 0.54 | 0.28 | 0.00 |
| 8 | Google Glass Enterprise Edition 2 (2019) | 0 | 0.75 | 0.50 | 0.38 | 0.25 | 0.05 | 1.00 | 1.00 | 0.50 | 0.00 |
| 9 | Kopin Golden-i Infinity Smart Screen (2018) | 0 | 0.42 | 0.81 | 0.00 | 0.00 | 0.19 | 0.26 | 1.00 | 0.56 | 0.00 |
| 10 | RealWear HMT-1Z1 (2018) | 0 | 0.83 | 1.00 | 0.25 | 0.13 | 0.21 | 0.25 | 0.11 | 0.08 | 0.00 |
| 11 | Kopin Solos Smart Glasses (2018) | 0 | 0.42 | 0.31 | 0.00 | 0.00 | 0.10 | 0.13 | 0.71 | 1.00 | 0.00 |
| 12 | Everysight Raptor (2017) | 0 | 0.67 | 0.81 | 0.25 | 0.25 | 0.13 | 0.94 | 0.47 | 0.77 | 0.00 |
| 13 | Epson Moverio BT-350 Smart Glasses (2017) | 0 | 0.58 | 0.31 | 0.25 | 0.25 | 0.18 | 0.29 | 0.36 | 0.45 | 0.00 |
| 14 | RealWear HMT-1 (2017) | 0 | 0.83 | 1.00 | 0.25 | 0.13 | 0.20 | 0.25 | 0.12 | 0.20 | 0.00 |
| 15 | Glassup F4 visor (2017) | 0 | 0.58 | 0.31 | 0.13 | 0.13 | 0.24 | 0.28 | 0.18 | 0.23 | 0.00 |
| 16 | ODG R-9 (2017) | 0 | 0.42 | 0.50 | 0.75 | 1.00 | 0.08 | 0.63 | 0.25 | 0.28 | 0.00 |

## 4.1 Time machine function

The time machine function enables a 3D representation of the past to be saved and restored to be displayed simultaneously with the immediate environment. The AR device can track users' positions in space using the spatial mapping function to place stable and accurate holograms despite the users' movement. Therefore, even if the perspective changes the AR device can recognize proper positioning. The following procedure is to fulfill the time machine function. The spatial mapping of the real environment is defined at time t as

$$\{\mathbf{I}_n\}_t = (x, y, z, t)^T, n \in [1, N] \tag{2}$$

where $n$ represents the $n^{\text{th}}$ mesh element of the real environment; $N$ is the total number of the mesh elements in this environment; $\mathbf{I}_n$ is the spatial mapping consisting of $n$ mesh elements; $t$ represents the timestamp of the spatial mapping; and $(x, y, z)$ are the three-dimensional position coordinates.

Figure 2 explains the time machine function with the movement of a lamp. The lamp labeled with a red frame is a real object (i.e., a physical lamp in the room), and the one with a blue frame is virtual (i.e., simulated meshes of a lamp). In Figure 2(a), the environment is mapped, and the meshes of the lamp in the position $P_1$ are saved as $\{\mathbf{I}_n\}_{t_1}$. In Figure 2(b), the lamp has been moved from position $P_1$ to $P_2$, and $\{\mathbf{I}_n\}_{t_1}$ is restored, i.e., the virtual lamp at position $P_1$. In summary, with the time machine function, the users can observe the saved images from the past and the real object in the environment at the same time.

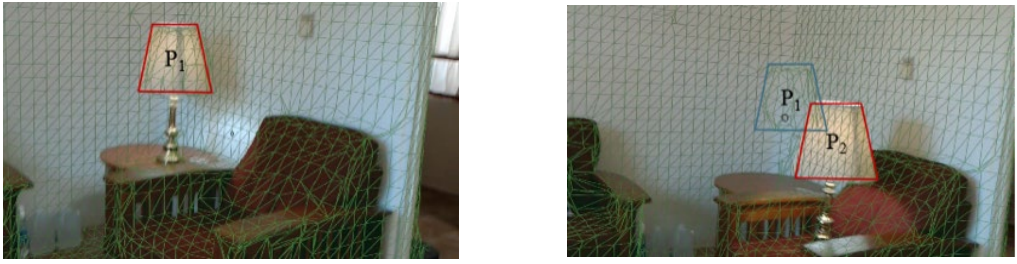

(a) Real image $\{\mathbf{I}_n\}_{t_1}$        (b) Real image $\{\mathbf{I}_n\}_{t_2}$ and virtual image $\{\mathbf{I}_n\}_{t_1}$

**Figure 2** Illustration of the time machine function

The time machine function utilizes Microsoft's Mixed Reality Toolkit (MRTK) (2020) mesh observer to add a collection of meshes. The meshes represent the geometry of the nearby real-world environment in the Unity scene. The MRTK mesh observer gets real-time environment data from the HoloLens Holographic Processing Unit (HPU). Figure 3 summarizes the saving and loading procedure in the TMM. Figure 4 shows a more detailed description of the time machine design frame.

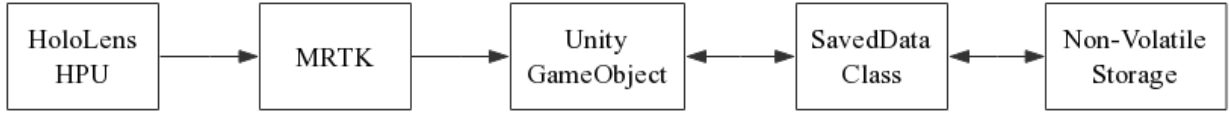

**Figure 3** Time machine function components

The mesh vertices, triangles, and time are saved to the non-volatile memory as an XML file for later access. This process is performed by a custom class to make a deep copy of the vertices and triangles of the meshes from the MRTK mesh observer. A deep copy is required because the mesh data is read-only, and Unity components cannot be directly serialized. The custom class then adds a timestamp to the deep copied meshes, copies the mesh positions relative to the observer, and serializes the System.Xml.Serialization component (Joshi 2017) from the .NET Runtime Library (Barnett and Schulte 2003).

To load a saved mesh, a Unity GameObject is created with four components attached, i.e., MeshFilter, MeshRenderer, MeshCollider, and TimeMachineData. The MeshFilter, MeshRenderer, and MeshCollider are components provided by the Unity game engine to enable the rendering and collision of meshes. The TimeMachineData is a custom-defined component. The saved XML file is then deserialized into an int array that defines the mesh triangles, a 3-point vector array that defines the mesh vertices, and a timestamp. The triangles and vertices are then copied to the MeshFilter; the timestamp is assigned to the TimeMachineData, and the new GameObject's position is set to the saved position.

## 4.2 Measure function

The 3D transformation during from time $t_1$ to $t_2$ consists of a 3D rotation $\Delta\mathbf{I}_{\mathrm{Rot}}$ and a 3D translation $\Delta\mathbf{I}_{\mathrm{Tra}}$. Combining $\Delta\mathbf{I}_{\mathrm{Rot}}$ and $\Delta\mathbf{I}_{\mathrm{Tra}}$ in a transformation matrix $\Delta\mathbf{I}$, then $\Delta\mathbf{I}$ is a rigid motion.

$$\Delta \mathbf{I} = \begin{bmatrix} \Delta \mathbf{I}_{\text{Rot}} & \Delta \mathbf{I}_{\text{Tra}} \\ \mathbf{0} & \mathbf{1} \end{bmatrix} \tag{3}$$

$$\{\mathbf{I}_n\}_{t_2} = \Delta \mathbf{I}_{t_1 \to t_2} \{\mathbf{I}_n\}_{t_1}, n \in [1, N] \tag{4}$$

The measure function is to measure $\Delta \mathbf{I}$, especially $\Delta \mathbf{I}_{\text{Tra}}$, as shown in Equation (3). The programming of the measure function is based on the spatial mapping that recognizes the surrounding environment. The authors use a script consisting of the Measure Manager, Input Manager, and Spatial Mapping components to conduct the measurement, as shown in Figure 5. The Measure Manager manages all the measurements in the Unity platform, and the Input Manager handles the input according to the users' sight and gesture. Specifically, the AR user performs an air tap gesture to place pins in the desired locations of measurement. Figure 6 shows a more detailed description of the measure function design frame.

The Spatial Mapping component consists of three scripts, Spatial Mapping Observer, Spatial Mapping Manager, and Object Surface Observer. The Spatial Mapping Observer script is to manage the Surface Observer component to conduct an environmental scan. The Object Surface Observer component scans the surrounding environment and loads the meshed objects. The Spatial Mapping Manager script has four functions: (i) restore the Surface Observer component, (ii) store the mesh data obtained by the Spatial Mapping component, (iii) shut down the Surface Observer component, and (iv) update the Spatial Mapping component in real-time while TMM is loading the mesh data.

### 4.3 Measurement over time

The authors combined the two primary functions of the TMM by leveraging Unity's layer system and a custom class to store time information on loaded GameObjects. Full meshes are assigned to the same layer as the meshes from the Spatial Observer component so that ray casts from the InputHandler (Figure 6) collide with the MeshCollider component. The InputHandler component checks the TimeMachineData component on any hit objects, allowing measurement over time by comparing the saved time of the hit meshes. If the InputHandler does not find a TimeMachineData component, it uses the current time to compare with the other meshes. Figure 7 shows a more detailed description of the measurement over time function design frame.

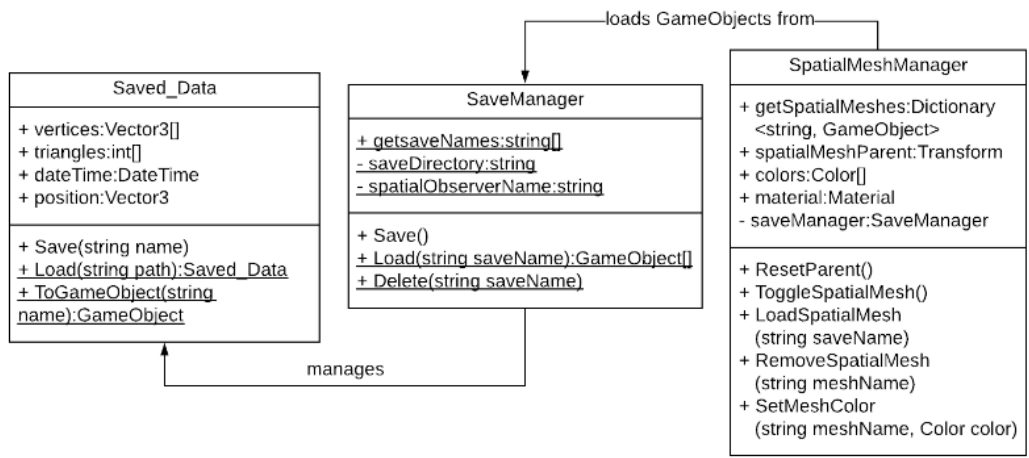

**Figure 4** Time machine function flowchart

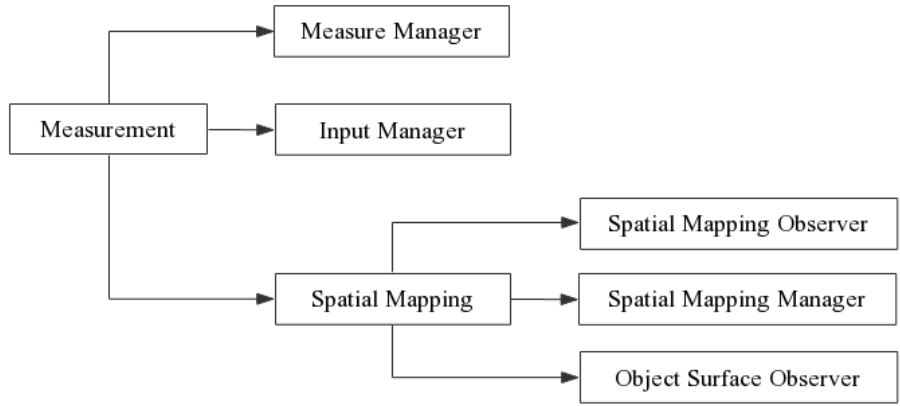

**Figure 5** Measure function components

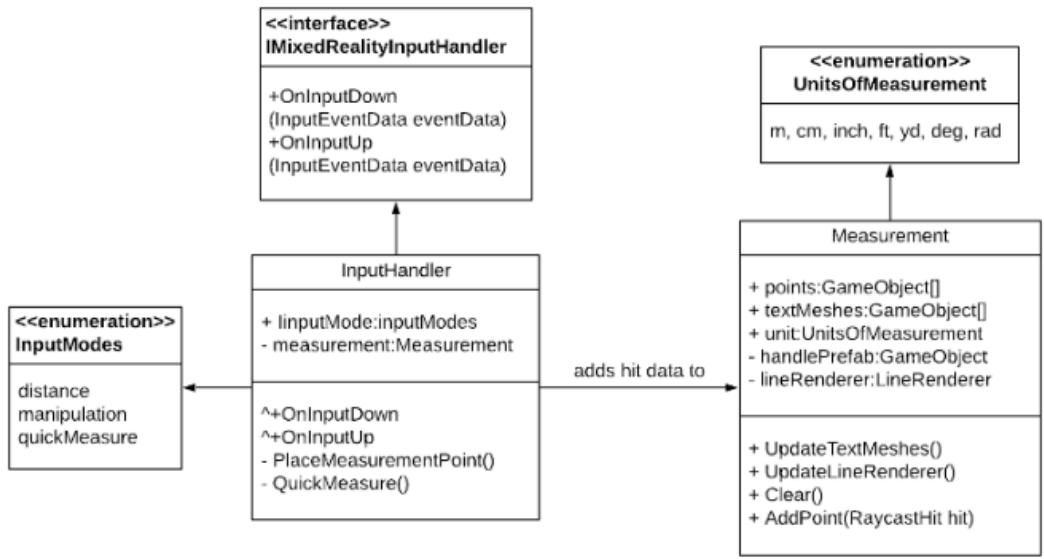

**Figure 6** Measure function flowchart

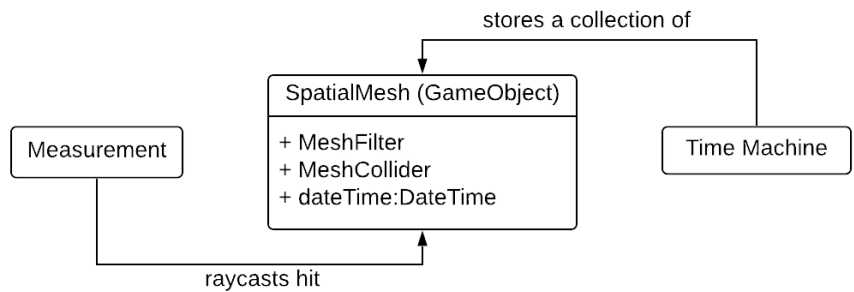

**Figure 7** Measurement over time flowchart

## 5 HUMAN-DEVICE INTERFACE

This section presents the human-device interface and illustrates the functions of different modules. Figure 8 shows the full view of the TMM human-device interface. The two modules in the TMM interface are the time machine menu and the measure menu. These correspond to their respective functions. The interface is developed with human factors considered. Each button is clearly labeled and the functions work simple enough for an inexperienced AR user. Figure 9 is a schematic diagram illustrating the interface during the implementation. The following subsections show a detailed explanation of the functions and instructions of each module.

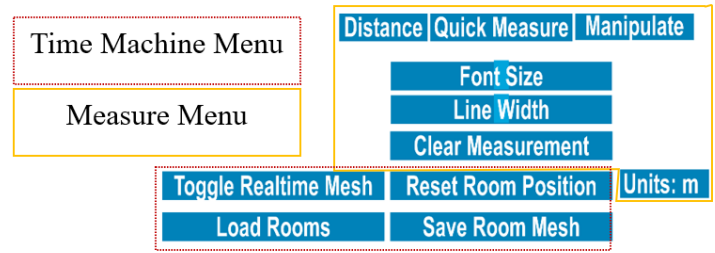

**Figure 8** The TMM human-device interface

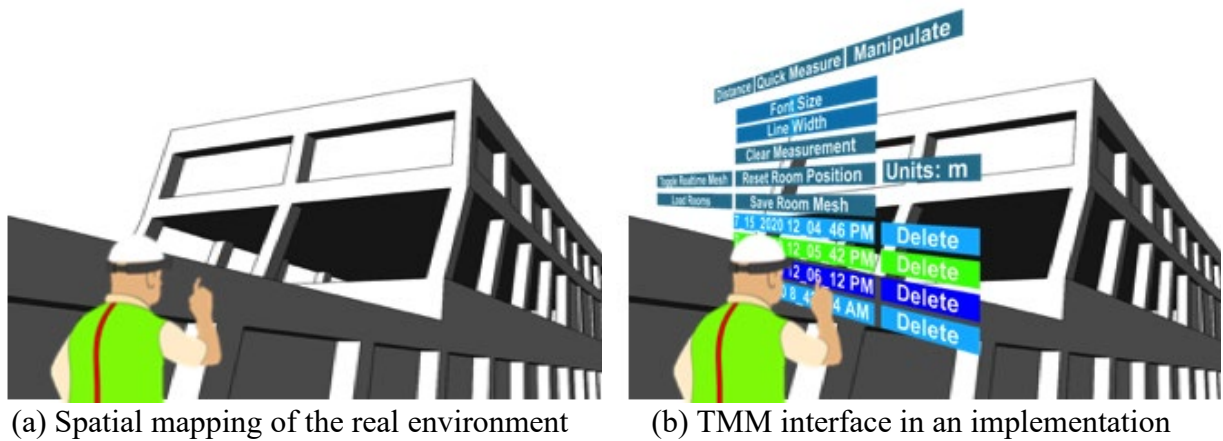

(a) Spatial mapping of the real environment          (b) TMM interface in an implementation

**Figure 9** Measuring Movements Across Time

## 5.1 Interface menus

The measure menu contains the following keys: Distance, Quick Measure, Manipulate, Font Size, Line Width, Clear Measurement, and Units. The time machine menu includes the following functions: Toggle Real-time Mesh, Reset Room Position, Load Rooms, and Save Room Mesh. Table 2 summarizes the functions of each key in detail. The authors name the keys in the interface according to their operations and show directly in the menu to decrease cognition load.

## 5.2 Control gestures

The user interfaces with the menus and keys of the TMM using hand gestures and their gaze following this sequence: (i) user gazes at the hologram meant to be clicked; (ii) user points the index finger straight up toward the ceiling; and (iii) user air taps the key intended to be accessed by first lowering the finger and then quickly raising it. Figure 10 summarizes the control gesture of the air tap.

**Table 2** The function of each key in the interface

| # | Menu | Key | Function |
|---|------|-----|----------|
| 1 | | Toggle Real-time Mesh | Scan the surroundings |
| 2 | Time Machine | Reset Room Position | Reset the environment and re-scan the surroundings |
| 3 | | Load Rooms | load a pre-scanned environment |
| 4 | | Save Room Mesh | Save the mapping of the current environment |
| 5 | | Distance | Measure distance between two points |
| 6 | | Quick Measure | Measure with default settings |
| 7 | | Manipulate | Downsize the virtual environments |
| 8 | Measure | Font Size | Modify the measurement font size |
| 9 | | Line Width | Modify the measurement line width |
| 10 | | Clear Measurement | Clear all measurements |
| 11 | | Units | Show the measurement units |

To conduct a measurement, a user first points to a node as the starting point with the index finger of their right hand, performing an air tap to place a pin. Next, they point to another node with a different timestamp endpoint (i.e., from a saved and loaded image) and place a second pin. The interface then shows the distance between the start and endpoints automatically. Multiple measurements can be conducted in the same space where the measure function displays the distance between each successive pin placement. Measurements can also be cleared for new measurements and the pins can be manually moved after they have been placed if the user desires a different or more exact location.

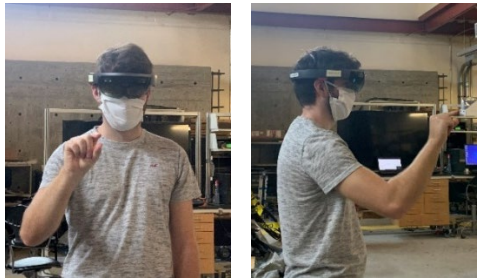

(a) Front image        (b) Profile image

**Figure 10** Air tap gestures to control the TMM by user

## 5.3 Color

TMM can simultaneously save a maximum of six environment meshes with different timestamps. Therefore, there could be a maximum of six virtual environments displayed overlapping onto the reality at a given time. The saved meshes with different timestamps are shown in different colors to decrease cognitive load for the users. An inspection site can have low visibility, and with this consideration the authors select opposite colors with high saturation for meshes in a sequence. Two contiguous meshes are restored with opposite colors according to the color wheel to generate high contrast and improve the visualizations of the variations in time (Cohen-Or et al. 2006). With the contrasting colors selected, the measurement procedure between different meshes will be less time-consuming. The colors implemented in the TMM interface are cyan, orange, lime, red, blue, and magenta. Table 3 shows the names, RBG values, and a palette of the selected colors.

**Table 3** Colors for the saved meshes in the TMM interface

| # | Color Name | RBG | Color Palette |
|---|---|---|---|
| 1 | Cyan | (0, 255, 255) | 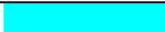 |
| 2 | Lime | (0, 255, 0) | 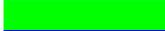 |
| 3 | Blue | (0, 0, 255) | 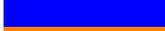 |
| 4 | Orange | (255, 128, 0) | 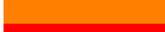 |
| 5 | Red | (255, 0, 0) | 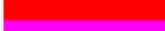 |
| 6 | Magenta | (255, 0, 255) | 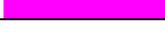 |

## 5.4 Special functionalities

Three functionalities are designed in the interface to help the users implement the TMM application. These are timestamp, downsizing, and automatic font scaling. The following subsections provide detailed explanations of these three functionalities.

### 5.4.1 Timestamp

Displaying timestamps helps distinguish how the environment has been changing over time. Without the timestamps, it would be difficult for the users to tell the measurement duration and it is valuable to document the time and date that prior captures were taken. Timestamps of the saved meshes are automatically loaded and displayed on the TMM interface, where each key has a date and timestamp with a color that matches the corresponding mesh. The demonstration of the TMM section shows examples of the timestamp feature.

### 5.4.2 Downsizing

The downsizing feature of the application allows the user to manipulate the size and angle of one capture or overlaid captures. This feature is necessary for easy viewing of the saved environment, enabling the users to inspect the surroundings quickly. It also allows the users to manipulate the view to take measurements and draw conclusions more quickly. Users can deploy the downsizing feature by air tapping the 'Manipulate' key (right upper corner in Figure 8) and then pinching with both hands. Users can move both hands closer to downsize the figure and change either hand's position to alter the angle. The demonstration of the TMM section shows an example of the downsizing feature.

### 5.4.3 Automatic font scaling

Both the font size and line width of the measurement results can be adjusted automatically or manually. Users can alter the font size and line width by the sliders on the 'Font Size' and 'Line Width' keys, as shown in Figure 8. The measurement text is rendered in 3D space and scaled relative to the user's distance to maintain font readability with different lengths. As shown in Figure 11, the font size for the measurements automatically keeps the same despite measuring distance.

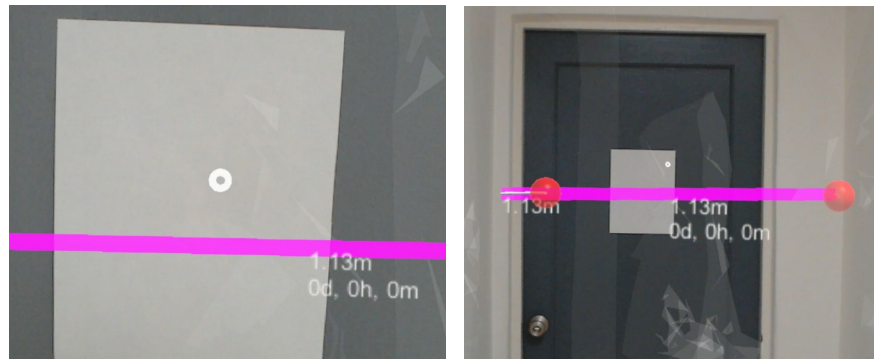

(a) Font at 0.5m                    (b) Font at 2.0m

**Figure 11** Automatic font scaling function

## 6 TMM EXPERIMENT

All the TMM application functions should be pre-verified before its implementation in a real inspection site. This experiment is to verify the accuracy of the measurement over time function and show its convenience and efficiency over the traditional measuring method.

### 6.1 Experiment description

Researchers conducted one experiment to verify the reliability of the TMM in comparison with traditional measurement methods. To this end, researchers moved one piece of furniture, a cube storage container, and measured the movement by both a traditional method with a tape measurer and TMM. Researchers marked the floor with the original and final position to compare the two methods (Figure 12(a)).

### 6.2 Measurements

The researchers validated the TMM by conducting two measurements simultaneously and comparing them. The two different measurement approaches are described herein, for comparison purposes:

(a) Figure 12(b) and (c) show the traditional tape measurement sequence: first, researchers marked the starting position of the object that would serve as a reference about its past position; then, after the movement, a mark was added that would serve as the present position of interest; finally, the researchers used a tape measure to measure the distance between the marks representing past and present positions, which was the movement across time.

(b) Figure 12(d) shows the measurement using the TMM: the user can collect the change on position without marks on the ground. The TMM overlapped the virtual objects from both the past and the present time of the movement simultaneously, and the user measured the movement of the object across time. Additionally, the user can measure more than one movement across time, enabling multiple comparisons across time with small effort.

## 6.3 Experimental results

The result by the traditional method is 0.91m, and by TMM is 0.93m. The measuring difference is about 2% between the two methods. The measuring difference is induced by (i) the spatial meshing precision; and (ii) the accuracy of the user's placement. It can be inferred that human error may cause some error, where the user may not have placed the AR pins perfectly at the centered of the marks on the floor. The difference of 2% is accurate enough for a real-time immediate environmental observation.

The traditional method takes about 5 min to set the markers and conduct the measurement. In comparison, the TMM uses less than 1 min, according to the time stamp in Figure 12(d). Setting markers is inconvenient, time-consuming, sometimes even not applicable during an inspection.

## 7 TMM DEMONSTRATION EXPERIMENT

After verifying the reliability of the TMM, the authors design a demonstration to show a complete process of TMM implementation.

### 7.1 Experiment objective

This demonstration simulates an environment in which inspectors can observe and measure changes across time. This demonstration shows the features of the TMM application by executing the same steps as an inspection task, capturing, and overlaying separate environments to measure the changes between captures over time. The authors conduct a demonstration experiment in a residential house to test the TMM in a realistic scenario where results can have the highest impact for inspections in the future.

### 7.2 Experiment procedure

There are seven steps in the demonstration:
(a) Nominate and save the initial environment when the inspectors enter the site to verify the timestamp function.
(b) Move the container to simulate the environmental change.
(c) Restore the saved initial environment to verify the time machine function.
(d) Measure over time using the measure function and save the current environment, verifying the measurement over time function.
(e) Move the container again to simulate a continuous environmental change.
(f) Restore the saved images and measure over time to verify the robustness of the measurement over time function.
(g) Manipulate the virtual images to verify the downsizing function.

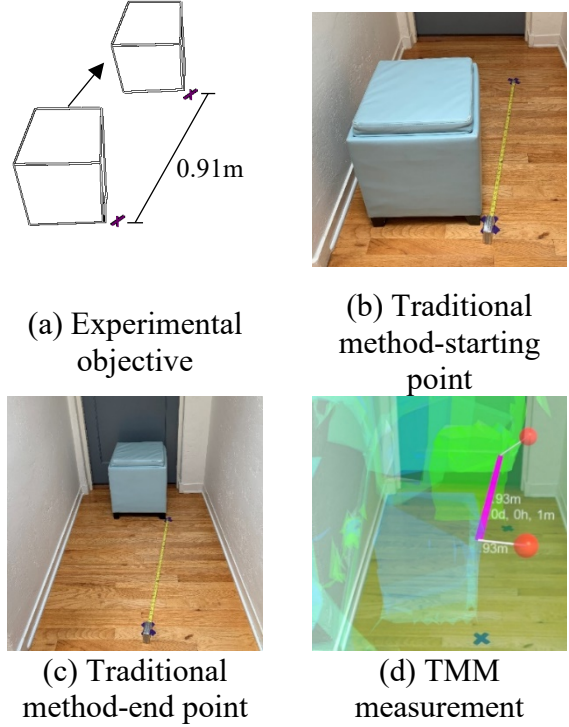

<div align="center">

(a) Experimental objective      (b) Traditional method-starting point

(c) Traditional method-end point      (d) TMM measurement

**Figure 12** TMM verification experiment

</div>

### 7.3 Experiment results

The following are the demonstration results:

(a) The authors nominate the time of the initial environment as $t_1$. The environment at time $t_1$ is saved with a timestamp and appointed as $\{\mathbf{I}_n\}_{t_1}$ (cyan color), as shown in Figure 13(a).

(b) The cube container is moved from the initial location to a new location during the time $t_1$ to $t_2$. The environment at time $t_2$ is saved with a timestamp and nominated as $\{\mathbf{I}_n\}_{t_2}$ (lime color).

(c) At time $t_2$, the saved $\{\mathbf{I}_n\}_{t_1}$ is restored as a virtual image. Both the real $\{\mathbf{I}_n\}_{t_2}$ and virtual $\{\mathbf{I}_n\}_{t_1}$ are shown to the user at the same time, as shown in Figure 13(b). The users can maneuver the color-coded environments to observe the container movement and verify the capture capabilities of the TMM.

(d) At time $t_2$, measure $\Delta \mathbf{I}_{t_1 \to t_2}$ by air tapping the container in each capture. The interface will then show the distance between points pinned at two desired locations with the elapsed time between the captures. The moving distance in step (b) is 0.87m, as shown in Figure 13(d).

(e) Save the environment at time $t_2$ as $\{\mathbf{I}_n\}_{t_2}$ (lime color). Furthermore, move the cube storage again, as shown in Figure 13(d).

(f) At time $t_3$, restore both $\{\mathbf{I}_n\}_{t_1}$ and $\{\mathbf{I}_n\}_{t_2}$ as virtual images. Measure $\Delta \mathbf{I}_{t_1 \to t_3}$. The total moving distance is 1.8m, as shown in Figure 13(e).

(h) The authors verify the downsizing feature by shrinking the three collaborative environments. With the downsizing function, users can conveniently obtain a whole picture of the surroundings in which they are standing. Users can observe the relative position of the object in the environment. Additionally, the downsizing function can clearly show the observed object's movement trail, shown as the red line in Figure 13(f).

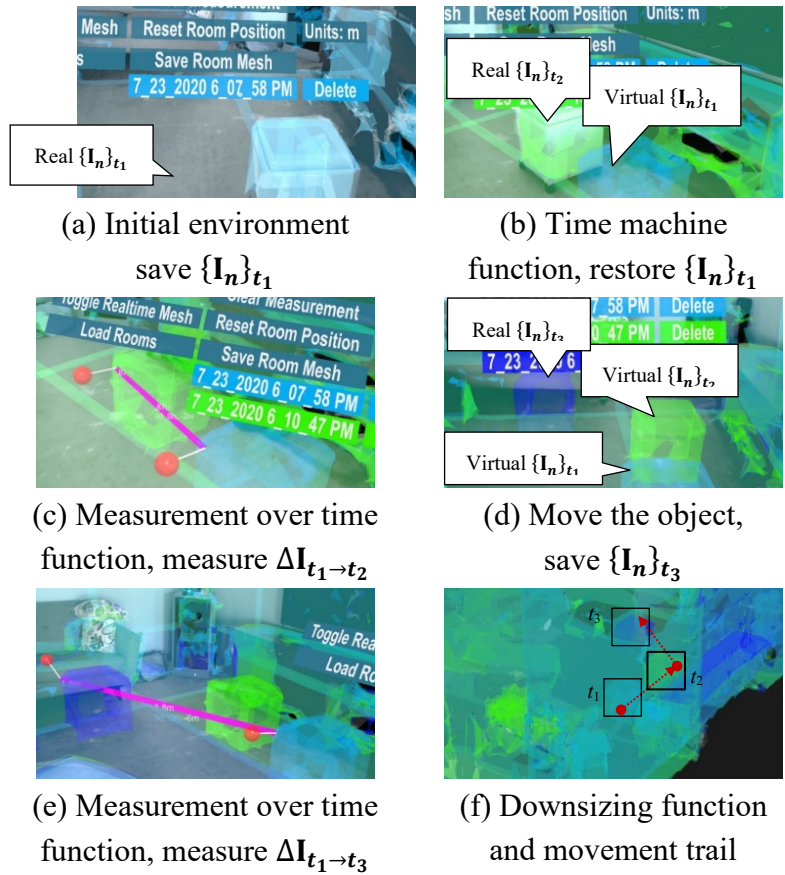

(a) Initial environment save $\{\mathbf{I}_n\}_{t_1}$

(b) Time machine function, restore $\{\mathbf{I}_n\}_{t_1}$

(c) Measurement over time function, measure $\Delta\mathbf{I}_{t_1\to t_2}$

(d) Move the object, save $\{\mathbf{I}_n\}_{t_3}$

(e) Measurement over time function, measure $\Delta\mathbf{I}_{t_1\to t_3}$

(f) Downsizing function and movement trail

**Figure 13** Demonstration of TMM implementation

## 8 CONCLUSIONS

SHM inspectors can benefit from a convenient method to observe and measure structural changes over extended time periods and in real-time to ensure safety following seismic events. Inspired by input from first responders for inspection priorities in buildings after disasters, the authors developed an AR application to measure the immediate environment movements over time. TMM has two primary functions, time machine and measure. With the time machine function, users can save and restore the nearby environment as virtual meshes. The virtual meshes can then be displayed overlapping onto the real environment. With the measure function, the users can measure the distance between the virtual and real meshes. Consequently, TMM enables inspectors to measure structural movement over time.

Besides the two primary functions, the researchers also design some practical functionalities for the TMM. First, the saved environments are stamped with different saturate and opposite colors to be easily distinguished. Then, the measurement fonts and line width can be automatically adjusted to minimize eye strain. Next, the virtual environment can be downsized for alternative viewing and inspection. Users can control the final version of TMM by gaze, hand gestures, and voice commands freeing both hands for inspection tasks. Finally, the researchers designed the human-device interface to be instinctively controlled to minimize cognitive load.

## ACKNOWLEDGMENTS

The financial support of this research is provided in part by the Air Force Research Laboratory (AFRL, Grant number FA9453-18-2-0022), and the New Mexico Consortium (NMC, Grant number 2RNA6.) The authors appreciate the discussion and feedback from the Department of Emergency Management in the City of Albuquerque. The conclusions of this research represent solely those of the authors.